\documentclass[prx,superscriptaddress,twocolumn,amsmath]{revtex4-2}

\usepackage[SquareTraceBrackets]{quantum}
\usepackage{graphicx,natbib,mathrsfs,amsmath,amsbsy,bm,optidef}
\usepackage{optidef}
\usepackage{siunitx}

\usepackage[lined,ruled,vlined]{algorithm2e}
\newcommand{\pname}[1]{\textsf{#1}}

\usepackage[caption=false]{subfig}

\usepackage{nameref}
\usepackage{varioref}

\usepackage{braket}

\usepackage[dvipsnames]{xcolor}
\definecolor{myblue}{named}{MidnightBlue}

\usepackage{tabularx}
\graphicspath{ {./Graphics/} }
\DeclareGraphicsExtensions{.pdf, .jpg, .eps, .svg, .png}
 
\usepackage[colorlinks=true,citecolor=myblue,linkcolor=myblue,urlcolor=myblue]{hyperref}

\begin{document}

\title{A privacy-preserving publicly verifiable quantum random number generator}


\author{Tanvirul Islam}
\affiliation{Centre for Quantum Technologies, National University of Singapore, 3 Science Drive 2, 117543 Singapore}

\author{Anindya Banerji}
\affiliation{Centre for Quantum Technologies, National University of Singapore, 3 Science Drive 2, 117543 Singapore}

\author{Chin Jia Boon}
\affiliation{Centre for Quantum Technologies, National University of Singapore, 3 Science Drive 2, 117543 Singapore}
\author{Wang Rui}
\affiliation{Centre for Quantum Technologies, National University of Singapore, 3 Science Drive 2, 117543 Singapore}

\author{Ayesha Reezwana}
\affiliation{Centre for Quantum Technologies, National University of Singapore, 3 Science Drive 2, 117543 Singapore}

\author{James A. Grieve}
\affiliation{Centre for Quantum Technologies, National University of Singapore, 3 Science Drive 2, 117543 Singapore}
\affiliation{Quantum Research Centre, Technology Innovation Institute, Abu Dhabi, United Arab Emirates}

\author{Rodrigo Piera}
\affiliation{Quantum Research Centre, Technology Innovation Institute, Abu Dhabi, United Arab Emirates}

\author{Alexander Ling}%
\affiliation{Centre for Quantum Technologies, National University of Singapore, 3 Science Drive 2, 117543 Singapore}
\affiliation{Department of Physics, National University of Singapore, Blk S12, 2 Science Drive 3, 117542 Singapore}

\begin{abstract}

    Verifying the quality of a random number generator involves performing computationally intensive statistical tests on large data sets commonly in the range of gigabytes. Limitations on computing power can restrict an end-user's ability to perform such verification. 
    There are also applications where the user needs to publicly demonstrate that the random bits they are using pass the statistical tests without the bits being revealed. 
    We report the implementation of an entanglement-based protocol that allows a third party to publicly perform statistical tests without compromising the privacy of the random bits. 
 
\end{abstract}

\maketitle

\section{Introduction}
\label{introduction}

\noindent

Generating random numbers that are private, secure, and have the statistical properties expected of a uniform randomness distribution is a crucial step for many computational tasks. For example, scientific simulations~\cite{hastings1970monte}, self-testing quantum systems~\cite{vsupic2020self}, randomized algorithms~\cite{rabin1980probabilistic,howes2007efficient}, machine learning~\cite{zhang2016survey}, cryptography~\cite{schindler2009random,bennett2020quantum}, lottery, gambling, public tenders, computer games, utilize random numbers during initialization of the systems or during operation. 
Pseudo-random number generators (PRNG) based on algorithms can have good statistical properties resembling a uniform source, but strong long-range correlations exist in the output that may undermine the applications~\cite{yuan2015randomness}, or introduce security loopholes.
This is because the seed to the PRNG is the only entropy in the system, and entropy cannot be increased by deterministic computation.
Quantum random number generators (QRNG)~\cite{ma2016quantum,herrero2017quantum} have been proposed as an alternative where entropy is extracted from a quantum mechanical process.    

All random number generators, however, face two common problems. First, the user may lack sufficient computational capacity to perform the statistical tests~\cite{l2007testu01,sonmez2008independence,luengo2021recommendations} needed to certify the quality of the randomness. Second, in public-facing applications, such as lottery or public tenders, the owner of the QRNG device may have to prove the statistical quality of the bits to public stakeholders before the bits are used. These problems demand a solution where a user may publicly test their random bits without revealing them.  
 
In a publicly testable random number generator~\cite{jacak2020quantum} multiple correlated streams of random bits are generated. A public tester performs arbitrary statistical tests on one of the bit streams to certify its randomness properties. This certifies the other output streams that are not shared with the tester. 

In this manuscript we report the implementation of a QRNG using only a polarization-entangled photon pair source, and linear optics. The implementation satisfies the conditions of secrecy, public testability.

\section{The protocol construction} 
\label{protocol}

A publicly verifiable QRNG should have the following properties. 

\begin{itemize}
    \item{\textbf{Property 1:}} The source of the entropy is of quantum origin. 
    \item{\textbf{Property 2:}} The quality of the random output is publicly verifiable without compromising the secrecy of the final output bits.
\end{itemize}

 In the following sections, we elaborate the steps of the protocol and demonstrate its implementation. 

\subsection{Publicly verifiable quantum random number generation protocol}

Property 1 is satisfied when an entanglement-based QRNG demonstrates that the source is producing a stream of  entangled states and the random output is generated from the outcome of projective measurements on these entangled qubits. Here, the entanglement can be verified using Bell inequalities~\cite{clauser1969proposed}.
In our implementation below we use the CHSH inequality to ensure that Property 1 is satisfied. 

A QRNG that produces a single stream of bits cannot be publicly verified without completely losing its secrecy. 
One needs a solution with at least two streams of bits, denoted $X_A$ and $X_B$, that are correlated in a way that publicly verifying the randomness of stream $X_A$ ensures the quality of the stream $X_B$. However, the protocol must ensure that their mutual information $I(X_A,X_B)=0$. When this is achieved, the bit stream $X_B$ can be securely used as a publicly verified private randomness. 

In our protocol, the QRNG produces three streams of random bits that are correlated. 
One of the bit streams is subjected to public randomness testing. As the streams are correlated this public randomness test verifies the quality of randomness in the other two streams that are not revealed. This satisfies Property 2.

To achieve Property 1 and 2, we prepare a tripartite entangled state, 
\begin{align} 
    \ket{\Phi_{ABC}}=\frac{1}{2}(\ket{000} - \ket{011}+\ket{101} - \ket{110}) \label{eq:tripartite_state}
\end{align}

This state exhibits the interesting property that performing a projective measurement in the computational basis on any one of the qubits projects the combined state of the other two qubits to either of two Bell states. As an example, if we measure qubit A in the computational basis the BC system is projected onto either Bell states, $\ket{\Phi^-_{BC}}$ or $\ket{\Psi^-_{BC}}$,  
\begin{align} \label{eq:tripartite_state_resolved}
    \ket{\Phi_{ABC}}=&\frac{1}{\sqrt{2}}\lbrace\ket{0}\left(\frac{\ket{00} - \ket{11}}{\sqrt{2}}\right) +\ket{1}\left(\frac{\ket{01} - \ket{10}}{\sqrt{2}}\right)\rbrace \\
    =&\frac{1}{\sqrt{2}}\lbrace\ket{0}\ket{\Phi^-_{BC}} +\ket{1}\ket{\Psi^-_{BC}}\rbrace 
\end{align}
 
Qubits prepared in a Bell state produce random outcomes when measured individually.
Monogamy of entanglement~\cite{terhal2004entanglement} ensures that this measurement outcome is not correlated to any outside system.  
Therefore, the outcome of the system $BC$ cannot be predicted even if one has access to the outcome of $A$.

Consider a single copy of the state~\eqref{eq:tripartite_state}. We perform a projective measurement in the computational basis on the three subsystems of the state. Let $x_A, x_B$ and $x_C$ denote the outcomes of projective measurement of the three subsystems, $A,B$ and $C$ in the computational basis. They can be considered as bit valued random variables taking their values with probabilities from Table~\ref{tab:probability_table}.

\begin{table}[h]
\begin{tabular}{cccc}
$p(x_A,x_B,x_C)$ & $x_A$ & $x_B$ & $x_C$ \\
1/4             & 0     & 0     & 0     \\
1/4             & 0     & 1     & 1     \\
1/4             & 1     & 0     & 1     \\
1/4             & 1     & 1     & 0    
\end{tabular}
\caption{Probability $p(x_A,x_B,x_C)$, of measurement outcomes $x_A$, $x_B$ and $x_C$ when each of the qubits A, B and C are subjected to projective measurement in the computational basis. If any one of the output columns is removed the remaining two columns shows uniform distribution of two bits, indicating they are mutually independent. Outcomes that are not presented in the table have probability 0.}
\label{tab:probability_table}
\end{table}

By construction of the state $\ket{\Phi_{ABC}}$ the outcomes always satisfy,
\begin{align} 
    x_A \oplus x_B \oplus x_C = 0 \label{eq:xor_condition}
\end{align}
where $\oplus$ is the addition modulo 2 operator. 

Table~\ref{tab:probability_table} shows that the marginal probability distribution for $x_A$ is, $p(x_A=1)=p(x_1=0)=1/2$. Also, $x_B$ and $x_C$ has similar marginal distribution.
Therefore, if we consider the each of the three bits individually then they have maximal Shannon entropy, 
\begin{align}
    H(x_A) = H(x_B) = H(x_C) = 1. \label{eq:equal_entropy}
\end{align}


From Table~\ref{tab:probability_table} we see that in the absence of knowledge of any one bit, the two other bits become completely uncorrelated with each other. That is, their marginal distribution factorises. Therefore, their mutual information is 0, 
\begin{align}
    I(x_A,x_B) = I(x_B,x_C)= I(x_C,x_A) = 0. \label{eq:mutual_info}
\end{align}

\begin{figure}
\includegraphics[scale=0.46]{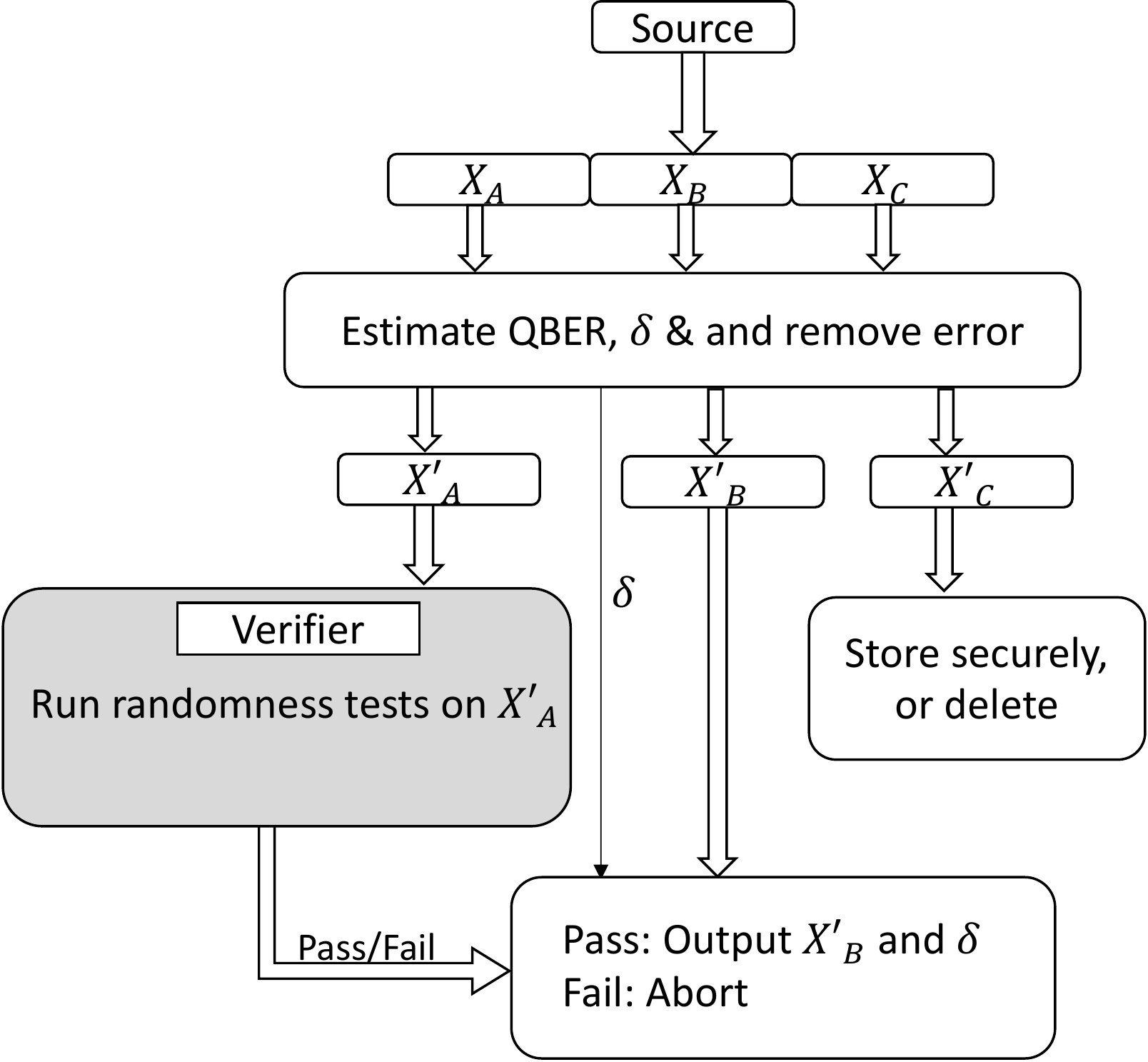}
  \caption{
  The QRNG outputs three correlated streams of random bits $X_A, X_B$ and $X_C$. Using them the quantum bit error rate (QBER), $\delta$ is estimated and the error triplet of bits are removed to generate $X'_A, X'_B$ and $X'_C$ . 
  After this, $X'_A$ is sent to public verifier $X'_C$ is stored securely or deleted. Verifier runs randomness tests on $X'_A$. If the test fails the protocol is aborted, else user outputs $X'_B$ and $\delta$. 
  }
  \label{fig:publicTesting}
\end{figure}

For random number generation, $n$ copies of the state $\ket{\Phi_{ABC}}$ prepared as in equation~\eqref{eq:tripartite_state} and each of the three parts of the state is measured in the computational basis. The outcomes are recorded in bit strings $X_A, X_B$ and $X_C$ of lengths $n$.  From our discussion so far, we see that each of the bit strings valued random variable $X_A, X_B$ and $X_C$ takes the value from strings in $\{0,1\}^n$ uniformly at random. 
 
From the preparation, each copy of the state~\eqref{eq:tripartite_state} are independent. Therefore, the condition~\eqref{eq:mutual_info} ensures that the random variables $X_A, X_B$ and $X_C$ are pairwise mutually independent. That is,
\begin{align}
    I(X_A,X_B) = I(X_B,X_C)= I(X_C,X_A) = 0. \label{eq:mutual_info_string}
\end{align}
The string $X_A$ is provided to a public verifier that validates the string via statistical tests. 
If $X_A$ passes the randomness test, 
condition~\eqref{eq:equal_entropy} ensures the quality of randomness of $X_B$ and $X_C$.
As the verifier only has access to $X_A$, the condition~\eqref{eq:mutual_info_string} ensures that no information is leaked about $X_B$ or $X_C$. 
However, following Eq.~\eqref{eq:xor_condition} knowledge of any two bit strings would allow recovery of the third string. Therefore to satisfy Property 2, either $X_B$ or $X_C$ should remain inaccessible.

\begin{algorithm}
\SetAlgorithmName{Protocol}{protocol}{List of Protocols} 
	\LinesNumbered
	\DontPrintSemicolon
	\SetKwInOut{Input}{input}\SetKwInOut{Output}{output}
	\Input{$n$ copies of the state $\ket{\Phi_{ABC}}$ prepared as in equation~\eqref{eq:tripartite_state}.}
	\Output{Publicly verified private random bits and QBER, or Fail.}

\SetKwBlock{GEN}{User: Generation}{}	

\GEN{
	Measure each part of the state $\ket{\Phi_{ABC}}$ in computational basis and store the outcome of system $A$ in $x_A$, $B$ in $x_B$ and $C$ in $x_C$\; \label{step2}
	Perform step~\ref{step2} $n$ times to construct bit strings $X_A, X_B$ and $X_C$\.;
    Assign, $L = \{i: \text{s.t. } X_A[i]\oplus X_B[i]\oplus X_C[i] \neq 0 \} $, be the set of indices where the XOR condition~\eqref{eq:xor_condition} fails.\; 
    Assign, QBER $= \frac{|L|}{n}$.\;
    Create $X'_A, X'_B$ and $X'_C$ from  $X_A, X_B$ and $X_C$ respectively by removing elements with indices $i\in L$.\;
    Send $X'_A$ to public verifier.\;
}
\SetKwBlock{VERIFY}{Public Verifier}{}	
\VERIFY{
        Run randomness tests on $X'_A$. If the test fails output `Fail', else output `Pass'.\;
}
\SetKwBlock{OUTPUT}{User: Randomness output}{}	
\OUTPUT{
        Receive output from public verifier.\;
        If the verifier output is `Fail' then ouptut `Fail' and abort protocol, else, output $X'_B$ and QBER, and securely store or delete $X'_C$.\;
}
\caption{\pname{Publicly verifiable QRNG}} \label{prt:protocol1}
\end{algorithm}

Imperfections in any practical implementation will lead to equation~\eqref{eq:xor_condition} not being always satisfied. Counting the number of events that do not meet the XOR condition~\eqref{eq:xor_condition} 
provides the quantum bit error rate (QBER). Removing the erroneous triplet of outcomes from $X_A, X_B$ and $X_C$ gives $X'_A, X'_B$ and $X'_C$ each of length $m$ that satisfy,

\begin{align}
X'_A \oplus X'_B \oplus X'_C = 0,  
\end{align}
where $\oplus$ denotes bit-wise addition modulo-2 operation. 

At this point the user sends out $X'_A$ to the public verifier for statistical randomness testing. If the verification fails then the user will discard the data and start over. If the verification succeeds then the user uses $X'_B$ as private randomness and securely stores or deletes $X'_C$. The presence of positive QBER indicates information leakage to the environment. The user may use the QBER information to perform further randomness extraction to amplify the privacy (similar to privacy amplification~\cite{renner2005universally} in quantum key distribution).  

The workflow of the protocol is depicted in Figure~\ref{fig:publicTesting} and the detailed steps are listed in \pname{Protocol 1}.


\subsection{The experimental setup}
\label{experimental setup}

\begin{figure}[ht]
  \includegraphics[scale=0.75]{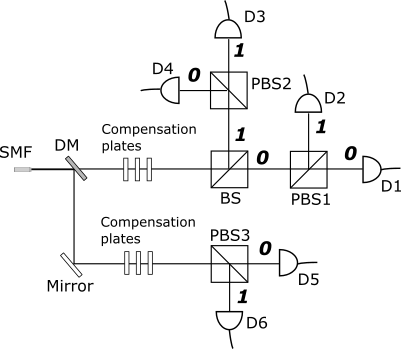}
  \caption{
  The detection setup.  The boldfaced numbers represent the bit values encoded by the path of photons and define the bit streams $X_A$, $X_B$ and $X_C$. Entangled photons are launched from a single mode fiber (SMF) and separated according to wavelengths by dichroic mirror (DM). 
  The polarization state of the photons in both paths are corrected by a stack of waveplates (Compensation plates). The output of the beam splitter (BS) generates $X_A$. Polarizing beam splitters PBS1 and PBS2 generate $X_B$. $X_C$ is generated by PBS3.
  }
  \label{fig:Detectors}
\end{figure}

The source of entangled photon pairs follows the design demonstrated in~\cite{lohrmann2020broadband} to produce photon pairs in the $\ket{\Phi^-}= \frac{1}{\sqrt{2}}(\ket{HH} - \ket{VV})$ Bell state.
Entangled photon pairs are emitted from a single mode fiber (SMF) to the detector setup (see Figure~\ref{fig:Detectors}) and the signal photons $(\lambda \approx 780nm)$ are separated from the idler photons $(\lambda \approx 842nm)$ by a dichroic mirror (DM). Stacks of quarter-half-quarter waveplates correct for the change in polarization state caused by the SMF birefringence.

The idler photons exit out of the two ports of the non-polarizing beam splitter (BS) with equal probability. This choice of paths defines the bit $x_A$. The idler photons are then projected into the $H/V$ basis by either polarizing beam splitters (PBS). The value of $x_B$ depends on the detection outcome at PBS1 or PBS2, and $x_c$ on the outcome at PBS3.

Due to the entanglement between the signal and idler photons, coincidence events are only expected to occur between the following detector pairs with equal probability: D1 and D5, D2 and D6, D3 and D5, D4 and D6. Together with $x_A$ determined from the choice of paths at the BS, the state in Eq. \ref{eq:tripartite_state} can be realised.
This is achieved by flipping the outcome labels in PBS2 compared to PBS1, which is equivalent to performing a local rotation of $\pi/2$ on path $1$. If path $0$ is taken at BS, then the detectors measure the state $\ket{\Phi^-_{BC}}$ and if path $1$ is taken, they measure $\ket{\Psi^-_{BC}}$. 

\subsubsection{Proof of Entanglement}

Generating a high fidelity Bell state  is crucial to prepare the state~\eqref{eq:tripartite_state} which preserves the secrecy of $X_B$ and $X_C$. Any QBER observed in the measurement outcome indicates the leakage of information to the environment and has to be taken care of in the privacy amplification step. 

\begin{figure}[ht]
  \includegraphics[scale=.55]{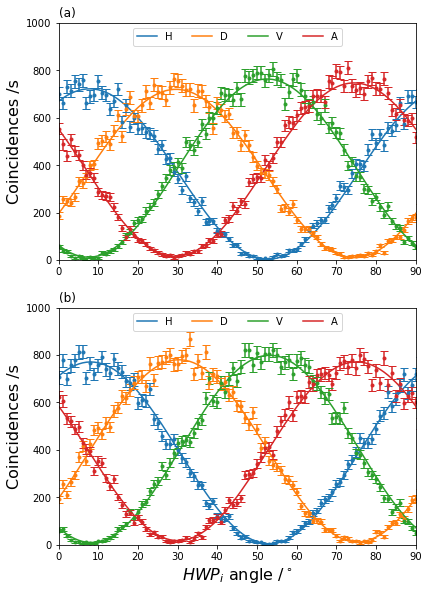}
  \caption{
    (a) Coincidences between  (D1,D2) and D5, with visibilities of $0.988 \pm 0.006, 0.971 \pm 0.009, 0.967 \pm 0.009, 0.96 \pm 0.01$ for the H, D, V and A bases respectively.
    (b) Coincidences between (D3,D4) and D5, with visibilities of $0.989 \pm 0.005, 0.969 \pm 0.005, 0.976 \pm 0.008, 0.96 \pm 0.01$ for the H, D, V and A bases respectively. 
    The visibilities for the coincidences between D1-4 and D6 (shown in supplementary material) are lower but are all above $0.93$. (Color online).
  }
  \label{fig:coincidences}
\end{figure}

In the experimental setup (Figure~\ref{fig:Detectors}, halfwave plates were placed before BS and PBS3 to measure the visibility curves (Figure \ref{fig:coincidences}) from which the CHSH~\cite{clauser1969proposed} values can be computed. The CHSH value for the state measured by systems (D1,D2) and (D5,D6) was $2.70 \pm 0.04$, while the value for the state measured by systems (D3,D4) and (D5,D6) was $2.72 \pm 0.04$. 

\subsubsection{Randomness Testing Results}
\begin{figure}[ht]
\includegraphics[scale=0.4]{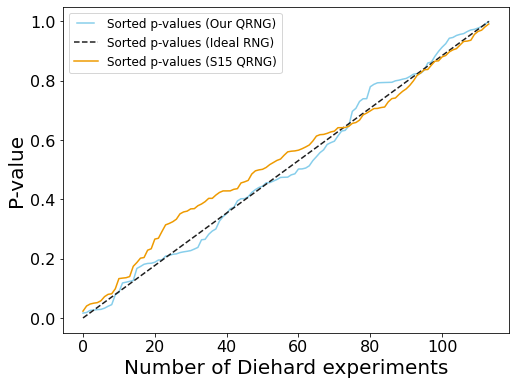}
  \caption{
   Sorted p-values of the statistical tests run by the diehearder randomness test suite. The black dashed line shows the expected ideal line. The blue curve shows one run of the test result on our data. The orange line shows the test result ran on same amount of data generated by the QRNG1~\cite{yicheng2020method} by S-Fifteen Instruments.
   The curves imply that our source shows close to ideal expected performance. 
   }
  \label{fig:KS_test}
\end{figure}

We perform the statistical randomness test suite `dieharder'~\cite{brown2004dieharder} on random numbers generated using our implementation of \pname{Protocol 1}.
This is to verify that the system is indeed generating good quality randomness.
Although a thorough verification of randomness would require larger size of data and significantly more computational resource, our limited test shows that the data is very close to an ideal randomness source. The system is compatible for running extensive tests by any third party certification process. Figure~\ref{fig:KS_test} shows a result for KS test~\cite{massey1951kolmogorov} that was ran on 1 MB of generated random bits. We run the same test on 1 MB of data from quantum random number generators by S-Fifteen Instruments and show it in the figure for comparison. 

\section{Discussion and future direction}
We have presented a QRNG source where the source stream can be subjected to public statistical randomness testing without compromising the secrecy of the final output bits. Any change in detector efficiencies can be locally checked before sending out for public randomness testing. 
This allow the user to remove statistical bias in the bit strings to avoid information leakage.
Along with robust miniaturized polarization entangled photon-pair sources, this setup can be built into a publicly verifiable QRNG source as a commercial off-the-shelf (COTS) product. Additionally, our entanglement based design can be extended to operated as a source device-independent~\cite{ma2016quantum} publicly verifiable auditable QRNG. 

\acknowledgments
This research is supported by the National Research Foundation, Singapore and A*STAR under its CQT Bridging Grant. We thank S-Fifteen Instruments for providing their QRNG data. 
\noindent

\bibliographystyle{ieeetr}
\bibliography{Biblio}
\end{document}